\documentclass[journal,10pt]{IEEEtran}

\usepackage{graphicx,epic,eepic,epsfig,amsmath,latexsym,amssymb,verbatim,hyperref}

\usepackage{verbatim}

\usepackage{theorem}
\newtheorem{definition}{Definition}[section]
\newtheorem{proposition}[definition]{Proposition}
\newtheorem{lemma}[definition]{Lemma}

\newtheorem{corollary}[definition]{Corollary}

\def\squareforqed{\hbox{\rlap{$\sqcap$}$\sqcup$}}
\def\qed{\ifmmode\squareforqed\else{\unskip\nobreak\hfil
\penalty50\hskip1em\null\nobreak\hfil\squareforqed
\parfillskip=0pt\finalhyphendemerits=0\endgraf}\fi}
\def\endenv{\ifmmode\;\else{\unskip\nobreak\hfil
\penalty50\hskip1em\null\nobreak\hfil\;
\parfillskip=0pt\finalhyphendemerits=0\endgraf}\fi}
\renewenvironment{proof}{\noindent \textbf{{Proof~} }}{\qed}

\newcommand{\exampleTitle}[1]{\textbf{(#1)}}

\newcommand{\proofComment}[1]{\exampleTitle{#1}}

\mathchardef\ordinarycolon\mathcode`\:
\mathcode`\:=\string"8000
\def\vcentcolon{\mathrel{\mathop\ordinarycolon}}
\begingroup \catcode`\:=\active
  \lowercase{\endgroup
  \let :\vcentcolon
  }

\newcommand{\nc}{\newcommand}
\nc{\rnc}{\renewcommand}
\nc{\beq}{\begin{equation}}
\nc{\eeq}{{\end{equation}}}
\nc{\beqa}{\begin{eqnarray}}
\nc{\eeqa}{\end{eqnarray}}
\nc{\lbar}[1]{\overline{#1}}
\nc{\bra}[1]{\langle#1|}
\nc{\ket}[1]{|#1\rangle}
\nc{\ketbra}[2]{|#1\rangle\!\langle#2|}
\nc{\braket}[2]{\langle#1|#2\rangle}
\nc{\proj}[1]{| #1\rangle\!\langle #1 |}
\nc{\avg}[1]{\langle#1\rangle}
\rnc{\max}{\operatorname{max}}
\nc{\Rank}{\operatorname{Rank}}
\nc{\smfrac}[2]{\mbox{$\frac{#1}{#2}$}}
\nc{\Tr}{\operatorname{Tr}}
\nc{\id}{\operatorname{id}}
\nc{\1}{\mathbb{I}}
\nc{\ox}{\otimes}
\nc{\dg}{\dagger}
\nc{\dn}{\downarrow}
\nc{\cA}{{\cal A}}
\nc{\cB}{{\cal B}}
\nc{\cC}{{\cal C}}
\nc{\cD}{{\cal D}}
\nc{\cE}{{\cal E}}
\nc{\cF}{{\cal F}}
\nc{\cG}{{\cal G}}
\nc{\cH}{{\cal H}}
\nc{\cI}{{\cal I}}
\nc{\cJ}{{\cal J}}
\nc{\cK}{{\cal K}}
\nc{\cL}{{\cal L}}
\nc{\cM}{{\cal M}}
\nc{\cN}{{\cal N}}
\nc{\cO}{{\cal O}}
\nc{\cP}{{\cal P}}
\nc{\cR}{{\cal R}}
\nc{\cS}{{\cal S}}
\nc{\cT}{{\cal T}}
\nc{\cX}{{\cal X}}
\nc{\cZ}{{\cal Z}}
\nc{\supp}{{\operatorname{supp}}}
\nc{\var}{\operatorname{var}}
\nc{\rar}{\rightarrow}
\nc{\lrar}{\longrightarrow}
\nc{\polylog}{\operatorname{polylog}}

\def\a{\alpha}
\def\b{\beta}
\def\g{\gamma}
\def\d{\delta}
\def\e{\epsilon}

\def\k{\kappa}
\def\l{\lambda}

\def\r{\rho}
\def\s{\sigma}

\def\ph{\varphi}

\def\ps{\psi}

\nc{\RR}{{{\mathbb R}}}
\nc{\CC}{{{\mathbb C}}}
\nc{\FF}{{{\mathbb F}}}
\nc{\NN}{{{\mathbb N}}}
\nc{\ZZ}{{{\mathbb Z}}}
\nc{\PP}{{{\mathbb P}}}
\nc{\QQ}{{{\mathbb Q}}}
\nc{\UU}{{{\mathbb U}}}
\nc{\EE}{{{\mathbb E}}}

\nc{\be}{\begin{equation}}
\nc{\ee}{{\end{equation}}}
\nc{\bea}{\begin{eqnarray}}
\nc{\eea}{\end{eqnarray}}
\nc{\<}{\langle}
\rnc{\>}{\rangle}
\nc{\Hom}[2]{\mbox{Hom}(\CC^{#1},\CC^{#2})}
\nc{\rU}{\mbox{U}}

\nc{\ob}[1]{#1}

\begin{document}
\title{Optimal superdense coding of entangled states}
\author{Anura Abeyesinghe, Patrick Hayden, Graeme Smith, and Andreas Winter \thanks{A. Abeyesinghe, P. Hayden and G. Smith are with the Institute for Quantum Information, Caltech 107--81, Pasadena, CA 91125, USA (email: anura@caltech.edu, patrick@cs.caltech.edu, graeme@caltech.edu)}\thanks{A. Winter is with the Department of Mathematics, University of Bristol,Bristol BS8 1UB, United Kingdom (email: a.j.winter@bris.ac.uk)}\thanks{AA, PH, and GS acknowledge the support of the US National
Science Foundation under grant no. EIA-0086038. 
PH acknowledges the support of the Sherman Fairchild Foundation. 
AW is grateful for the hospitality of the Caltech's Institute for 
Quantum Information, during a visit to which
part of the present work was done; he furthermore acknowledges support
by the EC project RESQ under contract no.~IST-2001-37559.
}}


\maketitle

\begin{abstract}
We present a one-shot method for preparing pure entangled states
between a sender and a receiver at a minimal cost of entanglement and
quantum communication. In the case of preparing unentangled states,
an earlier paper showed that a $2l$-qubit quantum state could be 
communicated to a receiver by physically 
transmitting only $l+o(l)$ qubits in addition to consuming
$l$ ebits of entanglement and some shared randomness. When the states
to be prepared are entangled, we find that there is a reduction
in the number of qubits that need to be transmitted, interpolating between
no communication at all for maximally entangled states and the earlier
two-for-one result of
the unentangled case, all without the use of any shared randomness.
We also present two applications of our result:  a direct proof 
of the achievability of the optimal 
superdense coding protocol for entangled states produced by a memoryless 
source, and a demonstration that the quantum identification capacity of
an ebit is two qubits.
\end{abstract}

\begin{keywords}
  concentration of measure, entanglement, identification, remote state preparation, superdense coding 
\end{keywords}


\section{Introduction} \label{sec:introduction}

A sender's power to communicate with a receiver is frequently enhanced 
if the two parties share entanglement.  The best-known example of 
this phenomenon is perhaps superdense coding \cite{BW92}, 
the communication of 
two classical bits of information by the transmission of  one
quantum bit and consumption of  one ebit. If the sender knows the identity 
of the state to be sent, superdense coding of \emph{quantum states} 
also becomes possible, 
with the result that, asymptotically, two qubits can be communicated 
by physically transmitting one qubit and consuming one bit of 
entanglement~\cite{HHL03,HLW04}.  In \cite{HHL03} it was furthermore 
shown that a sender (Alice) can asymptotically share a two qubit entangled 
state with a receiver (Bob) at the same qubit and ebit rate, along with 
the consumption of some shared randomness. That result, however, 
failed to exploit
one of the most basic observations about superdense coding: highly 
entangled states are much \emph{easier} to prepare than non-entangled 
states. Indeed, maximally entangled states can be prepared with no 
communication from the sender at all. 

In this paper, we construct a 
family of protocols that take advantage of this effect, finding that 
even partial entanglement in the state to be shared translates directly 
into a reduction in the amount of communication required.
Recall that every bipartite pure state can be written in the form
$\ket{\ph_{AB}} = \sum_i \sqrt{\l_i} \ket{e_i} \ket{f_i}$, where
$\braket{e_i}{e_j}=\braket{f_i}{f_j}=\d_{ij}$ and $\l_i \geq 0$~\cite{NC00}.
Since the numbers $\sqrt{\l_i}$, known as \emph{Schmidt coefficients},
are the only local invariants of $\ket{\ph_{AB}}$,
they entirely determine the nonlocal features of the state.
In the case of one-shot superdense coding, we find that it is the largest
Schmidt coefficient that plays a crucial role. More specifically,
we show how Alice can, with fidelity at least $1-\k$, share with Bob any 
pure state that has reduction on Bob's system of dimension $d_S$ and
maximum Schmidt coefficient $\sqrt{\l_{\max}}$ by
transmitting $\frac{1}{2}\log d_S{+}\frac{1}{2}\log \l_{\max}{+}O\left(\log(1/\k)\log\log d_S\right)$ 
qubits and consuming 
$\frac{1}{2}\log d_S{-}\frac{1}{2}\log \l_{\max}{+}O\left(\log(1/\k)\log\log d_S\right)$ ebits. We also
show that these rates are essentially optimal.

In the spirit of \cite{DHW03}, this new protocol can be viewed as the 
``father'' of the noiseless, visible
state communication protocols. Composing it with teleportation generates
an optimal \emph{remote state preparation}~\cite{L00,BDSSTW01} protocol. 
Applying it to the
preparation of states drawn from a memoryless source generates all the
optimal rate points of the triple cbit-qubit-ebit trade-off studied
in~\cite{AH03}, when combined with quantum-classical 
trade-off coding~\cite{DB01,HJW02}. 
An inspiration for the present work was Harrow's
alternative construction of optimal protocols in this memoryless setting
that made use of coherent classical communication~\cite{H04} and pre-existing
remote state preparation protocols~\cite{BHLSW03}. Harrow's techniques provided
strong circumstantial evidence that the protocol we present here 
should exist.

The rest of the paper is structured as follows. We begin, in 
Section \ref{sec:protocol}, by 
presenting the universal protocol for superdense coding of entangled
states and then prove its optimality, along with that of the associated
remote state preparation protocol, in Section \ref{sec:optimality}. 
Section \ref{sec:memoryless} contains an easy application
of typical subspace techniques to the task of developing an optimal
protocol for preparing states generated by a memoryless source. Section
\ref{sec:ID} provides another application 
of the protocol, this time to the theory
of identification~\cite{RY79,AD89}. Specifically, we show that the
quantum identification capacity of an ebit is two qubits.

\noindent {\bf Notation:} 
We use the following conventions throughout the paper.
$\log$ and $\exp$ are always taken base $2$. Unless
otherwise stated, a ``state'' can be pure or mixed.
The density operator
$\proj{\ph}$ of the pure state $\ket{\ph}$ will frequently be written simply
as $\ph$.  If $\ph_{AB}$ is a state on $A\ox B$, we refer to the reduced state
on $A$ as $\ph_A$. Sometimes we omit subscripts labelling subsystems,
in which case the largest subsystem on which the state
has been defined should be assumed: $\ph = \ph_{AB}$ in the
bipartite system $A\ox B$, for example.
A system we call $A$ will have a Hilbert space also called $A$ with a 
dimension $d_A$. $\UU(d)$ denotes the unitary group on $\CC^d$, 
and $\cB(\CC^d)$ the
set of linear transformations from $\CC^d$ to itself.  We write the fidelity 
between two states $\rho$ and $\sigma$ as 
$F(\rho,\sigma) = \|\sqrt{\rho}\sqrt{\sigma}\|_1^2$ and the
von Neumann entropy of a state $\r$ as $S(\r)=-\Tr \r \log \r$.


\section{The Universal Protocol} \label{sec:protocol}
To begin, suppose that Alice would like to share a maximally entangled state 
with Bob.  Clearly, this can be accomplished without any communication -- 
Alice need only perform operations on her half of a 
fixed maximally entangled state shared between them.  In particular, if 
$\ket{\psi}$ is an arbitrary maximally entangled state and we denote by
$\ket{\Phi_d} = \frac{1}{\sqrt{d}}\sum_{i=1}^d \ket{i} \ket{i}$
a fixed maximally entangled state, then
$\ket{\psi}$ can be expressed as
\begin{equation}
\ket{\psi} = V_{\psi} \ox \1_B \ket{\Phi_d}, 
\end{equation}
where $V_{\psi}$ is a unitary transformation of Alice's system
which depends on $\psi$.  This identity is 
equivalent to the following circuit diagram, in which time runs
from left to right:
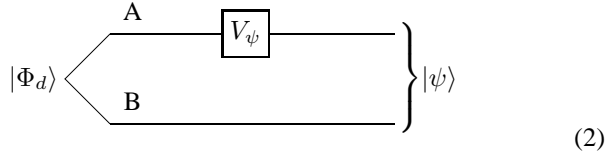
\bea
\setlength{\unitlength}{0.6mm}
\centering
\begin{picture}(90,35)
\put(-7,10){\makebox(10,10){$|\Phi_d\>$}}
\put(5,15){\line(1,1){10}}
\put(5,15){\line(1,-1){10}}
\put(15,25){\makebox(10,10){A}}
\put(15,5){\makebox(10,10){B}}
\put(15,25){\line(1,0){25}}
\put(15,5){\line(1,0){63}}
\put(40,20){\framebox(10,10){$V_{\psi}$}}
\put(50,25){\line(1,0){28}}
\put(80,10.3){\makebox(3,10){$\!\left.\rule{0pt}{5ex}\right\}$}}
\put(83,10){\makebox(10,10){$|\psi\>$}}
\end{picture}
\label{eq:maxEntangledProtocol}
\eea

Of course, in general, we would like to prepare an arbitrary state 
$\ket{\psi_{AB}}$ that may \emph{not} be maximally entangled, 
and to do so by using as few resources as possible.  
Our general method is as follows.
Alice and Bob initially 
share a fixed maximally entangled state $\ket{\Phi_{d_B}}$, to which
Alice applies
an isometry $V_\ps $. She then 
sends a subsystem $A_2$ of dimension $d_{A_2}$ to Bob, who 
applies a fixed unitary $U_{A_2B}^\dg$.  Alice's goal is to make 
$d_{A_2}$ as small as possible while still reliably preparing 
$\ket{\psi_{AB}}$.  The procedure can again be summarized with a
circuit diagram, although this time it is much less clear whether
there exist choices of the operations $V_\ps$ and $U_{A_2 B}$
that will do the job:
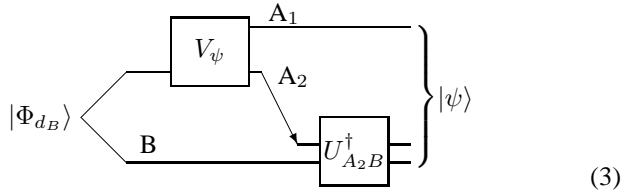
\bea
\setlength{\unitlength}{0.6mm}
\centering
\begin{picture}(90,42)
\put(-9,10){\makebox(10,10){$|\Phi_{d_B}\>$}}
\put(5,15){\line(1,1){10}}
\put(5,15){\line(1,-1){10}}
\put(45,33){\makebox(10,10){A$_1$}}
\put(47,19){\makebox(10,10){A$_2$}}
\put(15,04){\makebox(10,10){B}}
\put(15,25){\line(1,0){10}}
\put(15,05){\line(1,0){43}}
\put(25,22){\framebox(17,15){$V_{\psi}$}}
\put(42,25){\line(1,0){3}}
\put(42,35){\line(1,0){36}}
\put(45,25){\vector(1,-2){8}}
\put(53,09){\line(1,0){5}}
\put(80,02.7){\makebox(3,35){$\!\left.\rule{0pt}{7ex}\right\}$}}
\put(83,14){\makebox(10,10){$|\psi\>$}}

\put(73,09){\line(1,0){5}}
\put(73,05){\line(1,0){5}}

\put(58,0){\framebox(15,15){$U_{A_2B}^\dg$}}

\end{picture}
\label{eq:generalProtocol}
\eea

Circuit (\ref{eq:generalProtocol}) does provide a method for
preparing the state $\ket{\psi}$ as long as $\1 \ox U_{A_2B}\ket{\psi}$
is maximally entangled across the $A_1A_2 |B$ 
cut. (All such states are related by an operation on Alice's system alone.)
We will now use this observation, together with the fact that
high-dimensional states are generically highly 
entangled~\cite{L78,LP88,P93,FK94,S-R95,S96},
to construct a protocol that prepares
an arbitrary state with high fidelity. The precise statement about
the entanglement of generic states that we will need is the following lemma.

\begin{lemma} \label{lem:concentration}
Let $\ph$ be a state on $A \ox B$ proportional to a projector of rank $r$
and let $U_{AB} \in \UU(d_Ad_B)$ be chosen according to the Haar measure. 
Then, if $3 \leq d_B \leq d_A$,
\begin{eqnarray}
\Pr\big( S(\Tr_A U_{AB}\ph U_{AB}^\dg) < \log d_B - \a - \b \big)\nonumber \\
\leq 12 r \exp\left( -r d_A d_B \frac{\a^2 C}{(\log d_B)^2} \right),
\end{eqnarray}
where we may choose $C=(8\pi^2 \ln 2)^{-1}$,
and $\b = \frac{1}{\ln 2}\frac{d_B}{r d_A}$.
\end{lemma}

It generalizes the following lemma for rank-one $\ph$, 
which was proved in \cite{HLW04}.
\begin{lemma} \label{lem:badMeasurement}
Let $\ket{\ph}$ be chosen according to the Haar measure on $A \ox B$.
Then, if $3 \leq d_B \leq d_A$,
\begin{eqnarray}
\Pr\big( S(\ph_B) < \log d_B - \a - \b \big) \phantom{======} \nonumber \\
\leq \exp\left(- \left(d_B d_A-1 \right) \frac{\a^2 C}{(\log d_B)^2} \right),
\end{eqnarray}
where $C=(8\pi^2 \ln 2)^{-1}$ as before and
$\b = \frac{1}{\ln 2}\frac{d_B}{d_A}$.
\qed
\end{lemma}

\begin{proof}\proofComment{of Lemma \ref{lem:concentration}}
If we let $R$ be a space of dimension $r$ and
$\ket{\tau_{ABR}}$ be a uniformly distributed state on $A\ox B\ox R$, then
$\frac{1}{r}\Pi_{\tau_{AB}}$ is equal in distribution to 
$U_{AB}\ph U_{AB}^\dg$, where $\Pi_{\tau_{AB}}$ is 
the projector onto the support
of $\tau_{AB}$.  Let $\sigma$ denote the unitary transformation
$\sum_{j=0}^{r-1} \ketbra{e_{(j+1 \!\! \mod r)}}{e_j} + \1 - \Pi_{\tau_{AB}}$ 
that implements
a cyclic permutation on the eigenvectors $\{ \ket{e_j} \}$ of $\tau_{AB}$
corresponding to non-zero eigenvalues. (There are $r$ such eigenvalues 
except on a set of measure zero, which we will ignore.)
We then have
\begin{equation}\label{Eq:convexCombSupp}
\frac{1}{r}\Pi_{\tau_{AB}} 
=\frac{1}{r}\sum_{k=0}^{r-1} \sigma^k \tau_{AB}\sigma^{-k}.
\end{equation}
Eq.~(\ref{Eq:convexCombSupp}), together with the concavity of entropy, 
implies 
\begin{equation}
S\left(\frac{1}{r}\Tr_A \Pi_{\tau_{AB}}\right)
   \geq \frac{1}{r}\sum_{k=1}^{r} S(\Tr_A \sigma^k \tau_{AB}\sigma^{-k}),
\end{equation}
which in turn gives
\begin{eqnarray}
& &\!\!\!\!\!\!\!\!\!\!\!\!\!\!\!
    \Pr\left( S(\Tr_A U_{AB}\ph U_{AB}^\dg) < \log d_B - \a - \b \right) 
    \nonumber \\
&\leq& \Pr\Bigg( \frac{1}{r}\sum_{k=1}^{r}  S(\Tr_A \s^k \tau_{AB}\s^{-k}) 
   < \log d_B - \a - \b \Bigg) \nonumber \\
&\leq& r \Pr\bigl( S(\Tr_A \s^k \tau_{AB}\s^{-k}) < \log d_B - \a - \b \bigr) 
   \nonumber \\
&=& r \Pr\bigl( S(\tau_B) < \log d_B - \a - \b \bigr), 
  \label{Eq:equalEigsLastStep}
\end{eqnarray}
where the final step is a result of the unitary invariance of $\tau_{ABR}$. 
Applying Lemma \ref{lem:badMeasurement} 
to Eq.~(\ref{Eq:equalEigsLastStep}) with 
$A \rightarrow AR$ and  $B \rightarrow B$ reveals that 
\begin{eqnarray}
\Pr\left (S(\Tr_A U_{AB}\ph U_{AB}^\dg) < \log d_B - \a - \b \right) 
   \nonumber \\
   \phantom{==}
   \leq 12 r \exp\Big( - r d_B d_A \frac{\a^2 C}{4 (\log d_B)^2} \Big).
   \label{Eq:equalEigsResult}
\end{eqnarray}
\end{proof}

The idea behind the protocol is then simple: we will show that 
there exists a single unitary $U_{A_2 B}$ such
that $\1_{A_1} \ox U_{A_2 B}\ket{\psi_{A_1A_2B}}$ is almost maximally 
entangled across the $A_1 A_2| B$ cut for {\em all} states
$\ket{\psi_{A_1A_2B}}$ satisfying a bound on their Schmidt coefficients and
whose support on $A_2\ox B$ lies in 
a large subspace $S \subset A_2 \ox B$.
Since any such $\1_{A_1} \ox U_{A_2 B}\ket{\psi_{A_1A_2B}}$ is almost 
maximally entangled, we can then find an \emph{exactly} maximally entangled 
state which closely approximates it.
This state, in turn, can be prepared by the method of 
Circuit (\ref{eq:generalProtocol}). More formally, the following general
prescription can be made to succeed:

\medskip
\noindent
{\bf Protocol:} To send an arbitrary pure state with maximal Schmidt
coefficient $\leq \sqrt{\l_{\max}}$ and 
reduction of Bob's system to dimension $d_S$.
\begin{enumerate}
\item Alice and Bob share a maximally entangled state of
      $\log d_B = \smfrac{1}{2}( \log d_S - \log \l_{\max} ) + o(\log d_S)$ 
      ebits on their joint system $AB$.
\item Alice applies a local partial isometry $V_\ps$ with output on two
      subsystems $A_1$ and $A_2$. The size of $A_2$ is
      $\log d_{A_2} 
      = \smfrac{1}{2}( \log d_S + \log \l_{\max} ) + o(\log d_S)$.
\item Alice sends $A_2$ to Bob.
\item Bob applies $U_{A_2 B}^\dg$ followed by a projection onto $S$,
      which is embedded as a subspace of $A_2 B$.
\end{enumerate}
\begin{proposition} \label{prop:esdc}
Let $0 < \kappa \leq 1$. For sufficiently large $d_S$,and for
 $d_{A_2}$ and  $d_B$ as defined in the Protocol, there exists
choices of $V_\ps$ which depends on the input state
$\ket{\ps_{A_1 S}}$ and $U_{A_2 B}$ such that for all 
input states $\ket{\ps_{A_1 S}}$
with largest Schmidt coefficient $\leq \sqrt{\l_{\max}}$, 
the output of the protocol has
fidelity at least $1 - \kappa$ with $\ket{\ps_{A_1 S}}$.
\end{proposition}
\begin{proof}
Our method will be to show that if $U_{A_2B}$ is chosen according to 
the Haar measure, then the corresponding protocol has a nonzero probability 
over choices of $U_{A_2 B}$ of 
achieving high fidelity for all states that satisfy the 
restriction on their Schmidt coefficients, establishing the existence of a
particular $U_{A_2B}$ for which this is true. 

Now, to  ensure that the protocol succeeds on a given $\ket{\ps_{A_1 S}}$, 
we only need to ensure that $\1_{A_1}\ox U_{A_2B}\ket{\psi_{A_1 S}}$ is highly 
entangled across the $A_1 A_2 | B$ cut, which amounts to showing that
$S(\Tr_{A_2}U_{A_2 B}\psi_{S}U_{A_2 B}^\dg)$ is close to $\log d_B$.  This is 
exactly what Lemma \ref{lem:concentration} tells us is overwhelmingly likely
for an individual random
state $\ket{\ph_{A_1 S}}$ maximally entangled with a subspace $A_1'$ 
of $A_1$. By standard arguments, this will ensure that there exists
a unitary $U_{A_2B}$ such that $S(\Tr_{A_2}U_{A_2 B}\ph_{S}U_{A_2 B}^\dg)$ 
is close to $\log d_B$ for \emph{all} the states on $S$ maximally entangled
with $A_1'$. Majorization
can then be used to extend the argument to general states 
$\ket{\ps_{A_1 S}}$
with bounded largest Schmidt coefficient.

We begin by restricting to the case of states $\ket{\ph_{A_1' S}}$
maximally entangled between $S$ and a fixed 
subspace $A_1' \subseteq A_1$, with 
$d_{A_1'} = \lfloor 1/\l_{\max} \rfloor$.
Now, let $\cN^{\gamma}_{A_1' S}$ be a trace norm $\gamma$-net for such states.
It is possible to choose
$| \cN^{\gamma}_{A_1' S}| \leq (5/\g)^{2d_{A_1'} d_S}$ (See, for example,
\cite{HLSW03}). We will fix $\g$ later. 
By the definition of the net and the contractivity of the
trace norm under the partial trace,
for every maximally entangled state $\ket{\ph}$ on $A_1' \ox S$ 
there is a state 
$\ket{\tilde{\ph}} \in \cN^{\gamma}_{A_1' S}$ such that
\begin{eqnarray}
 \left\|  \Tr_{A_2}(U_{A_2B}\ph_S U_{A_2B}^\dg) 
         - \Tr_{A_2}(U_{A_2B}\tilde{\ph}_S U_{A_2B}^\dg) \right\|_1 
         \nonumber \\
                            \leq \|\ph - \tilde{\ph} \|_1 \leq \gamma,
\end{eqnarray}
which, by the Fannes inequality \cite{F73}, implies that
\begin{eqnarray}
&&\left| S\bigl( \Tr_{A_2}(U_{A_2B}\ph_S U_{A_2B}^\dg) \bigr) 
       - S\bigl( \Tr_{A_2}(U_{A_2B}\tilde{\ph}_S U_{A_2B}^\dg) \bigr) \right| 
       \nonumber \\
&& \quad\quad\quad\quad\quad\quad\quad\quad\quad\quad\quad\quad\quad 
   \leq \d + \eta(\g),
\end{eqnarray}
where $\d = \gamma \log d_B$ and $\eta(t) = -t\log t$ for $\g \leq 1/4$.  
Noting that all the states $\ket{\ps_{A_1 S}}$ have the same reduction on Bob, we have
\begin{eqnarray}
& &\!\!\!\!\!\!\!\!\!\!\!
    \Pr\Big( \inf_{\ket{\ph_{A_1' S}}} 
               S(\Tr_{A_2}U_{A_2B} \ph_{S}U_{A_2B}^\dg) \nonumber \\[-3.5ex] 
& &          \quad \quad \quad \quad \quad \quad 
               \Biggl.< \log d_B - \a - \b -\d - \eta(\gamma) \Big) 
              \label{eq:restrictedProb} \\[-1.5ex]
&\leq& |\cN^{\gamma}_{A_1' S}|
   \Pr\Bigl( \! S(\Tr_{A_2}U_{A_2B} \ph_{S}U_{A_2B}^\dg)
                                        \!<\! \log d_B - \!\a -\!\b \Bigr) 
               \nonumber \\
&\leq&  \left(\frac{5}{\gamma}\right)^{2 d_{A_1'}d_S} \!\! 4 d_{A_1'}
        \exp\left( - d_{A_1'} d_{A_2} d_B \frac{\a^2 C}{4 (\log d_B)^2} 
        \right)\!,
\label{eq:complicatedthing}
\end{eqnarray}
where 
$\b = d_B / (2 \ln 2 d_{A_1'}d_{A_2})$.
Choosing $\a = \b =: \e/4 \leq 1/4$,
$\g = \a^2/(4 \log d_B)$
and
\[
  d_S < d_{A_2} d_B \frac{\a^2 C}{8(\log d_B)^2\log(20\log d_B/\a^2)}-1,
\]
we find that the probability bound (\ref{eq:complicatedthing}) is less than 1.
For our choice of parameters, we have furthermore
$\a + \b + \d + \eta(\gamma) \leq 4\a = \e$, using
$\eta(x)\leq 2\sqrt{x}$ for $x\leq 1/4$.We have chosen parameters such that
$d_{A_2} =  d_B/(2\ln 2 \, \a d_{A_1'})$.

Moreover, relaxing the restriction on the input states now, 
suppose that $\ket{\ps}$ is any state on $A_1 \ox S$ satisfying the condition
$\| \ps_S \|_\infty \leq \l_{\max}$. Then any such $\ps_S$ is
majorized by any $\ph_S$ maximally entangled with $A_1'$,
so that
$\ps_S$ can be written as a convex combination 
$\sum_j p_j W_j \ph_S W_j^\dg$, where each 
$W_j$ is unitary~\cite{AU82}. 
It then follows from the concavity of the entropy that
\begin{eqnarray}
&& S\big(\Tr_{A_2} U_{A_2 B} \ps_S U_{A_2 B}^\dg\big) \nonumber \\
&& \phantom{===} 
\geq \min_j S\big( \Tr_{A_2} U_{A_2 B} W_j \ph_S W_j^\dg U_{A_2 B}^\dg\big).
   \phantom{===}
\end{eqnarray}
Therefore, the probability of Eq.~(\ref{eq:restrictedProb}) is actually an 
upper bound for 
\begin{equation}
\Pr\Big( 
    \underset{\ket{\ps_{A_1S}}}{\inf}S(\Tr_{A_2}U_{A_2B} \psi_{S}U_{A_2B}^\dg)
    < \log d_B - \e \Big).
\end{equation}
Thus, with our choice of parameters,
there is a unitary $U_{A_2 B}$ such that for 
all states $\ket{\psi}$ on $A_1\ox S$ 
satisfying the requirement that $\Tr_{A_1} \psi$ have 
eigenvalues $\leq \l_{\max}$, we have
\begin{equation}\label{Eq:entropEq}
  S(\ps'_B) \geq \log d_B - \e,
\end{equation}
introducing $\ket{\ps'} = (\1 \ox U_{A_2 B})\ket{\ps}$.
Since this can be rewritten as
$S\left( \ps_B'\|\1_B/d_B \right) = \log d_B - S(\ps'_B) \leq \e$,
it in turn implies~\cite{OhyaP93} that, for such states,
\begin{equation}
\| \psi'_B - \1_B/d_B\|_1 \leq \sqrt{2\ln 2 \e} =: \kappa, 
\label{Eq:traceEq} 
\end{equation}
and, therefore, that $F(\psi'_B,\1_B/d_B) \geq 1 - \kappa$. By
Uhlmann's theorem~\cite{Uhlmann76,Jozsa94}, there exists a purification 
$\Phi_\ps$ of $\1/d_B$ such
that $|\braket{\psi'}{\Phi_\ps}|^2 \geq 1 - \kappa$.
Starting from a fixed maximally entangled state $\ket{\Phi_0}$, 
$\ket{\Phi_\ps}$ can be prepared by Alice using a local operation 
$V_\ps$ on $A_1 A_2$ alone. 
Sending the system $A_2$ to Bob and having him perform
$U_{A_2 B}^\dg$ completes the protocol. The final state has fidelity at
least $1 - \kappa$ with $\ket{\ps}$.

We end with the accounting: the foregoing discussion
implies that we may choose
\begin{eqnarray}
\log d_{A_2} &=&  \frac{1}{2}\left(\log d_S + \log \l_{\max}\right)
                                                              \nonumber \\
             & &\phantom{===}  - O(\log\k) + O(\log\log d_S)  \nonumber \\
\log d_{B}   &=&  \frac{1}{2}\left(\log d_S - \log\l_{\max}\right)
                                                              \nonumber \\
             & &\phantom{===}  - O(\log\k) + O(\log\log d_S). \nonumber
\end{eqnarray}
\end{proof}

The main idea behind the proof, combining an exponential concentration
bound with discretization, has been used a number of times recently in
quantum information theory~\cite{HLSW03,BHLSW03,HHL03}. (It is, of course,
much older; see~\cite{MS86}.) If there is
a twist in the present application, it is illustrated in 
Eq.~(\ref{eq:complicatedthing}). Since $d_S$ is comparable in size
to $d_{A_2} d_B$, any prefactor significantly larger than 
$(5/\g)^{2d_{A_1'}d_S}$ would have caused the probability bound to fail.
Therefore, it was crucial to first restrict to states maximally
entangled between $A_1'$ and $S$, giving the manageable prefactor,
and then extend to general states and larger $A_1$ using majorization.

\section{Optimality of the protocol} \label{sec:optimality}

The communication and entanglement resources of Proposition \ref{prop:esdc}
are optimal up to terms of lower order than $\log d_S$ or
$\log\l_{\max}$: the amount of
quantum communication cannot be reduced, neither can the sum of the 
entanglement and quantum communication. (Entanglement alone can be
reduced at the cost of increasing the quantum communication.) We will
demonstrate the result in two steps. First we prove an optimality result
for the task of remotely preparing entangled quantum states using
entanglement and \emph{classical} communication. We then
show that by teleporting the quantum communication of our superdense
coding protocol for entangled states, we generate the optimal 
remote state preparation protocol, meaning the
original superdense coding protocol must have been optimal. 

\begin{proposition} \label{prop:ersp}
A remote state preparation protocol of fidelity $F\geq 1/2$
for all $d_S$-dimensional states
with maximum Schmidt coefficient $\leq \sqrt{\l_{\max}}$ must
make use of at least $\log d_S + \log\l_{\max} + \log F - 2$
cbits and $\log d_S - 18\sqrt{1-F}\log d_S - 2\eta\left(2\sqrt{1-F}\right)$
ebits, where $\eta(t)=-t\log t$.
\end{proposition}
\begin{proof}
Consider a remote state preparation protocol involving the 
transmission of exactly $\log K$ cbits which can, with fidelity $F$, 
prepare all $d_S$ dimensional states having maximum Schmidt coefficient 
$\sqrt{\l_{\max}}$.
We will show that causality essentially implies that 
$K$ must be roughly as large as $d_S\l_{\max}F$.

In particular, suppose Alice wants to send Bob a message
$i \in \left\{ 1, \dots, \lfloor\frac{d_S}{a}\rfloor \right\}$, with
$a=\lceil\frac{1}{\l_{\max}}\rceil$.
One way she can accomplish this is by preparing (a purification of) the state
$\sigma_i = \frac{1}{a}\sum_{k=1+a(i-1)}^{ai} \proj{k}$ on Bob's system,
with some fixed basis $\{\ket{k}\}$.
The remote state preparation protocol will produce a state $\rho_i$ for Bob
which will have a fidelity $F$ with the
intended state, $\sigma_i$. In order to decode
the message, Bob simply measures $\Pi_i = \sum_{k=1+a(i-1)}^{ai} \proj{k}$.
His probability of decoding the message Alice intended is
$\Tr(\rho_i \Pi_i) \geq F$.

Now, imagine that Alice and Bob use the same protocol,
with the modification that rather than Alice sending cbits,
Bob simply guesses which $j\in\{1,\dots ,K \}$ Alice 
would have sent.
The probability of Bob correctly identifying $i$ in this case is thus
at least $\frac{F}{K}$ --- he has a probability $\frac{1}{K}$ of correctly
guessing $j$ and, given a correct guess, a conditional probability $F$
of correctly identifying $i$.
However, since this protocol involves no forward communication from 
Alice to Bob, it can succeed with probability no greater 
than $\lfloor\frac{d_S}{a}\rfloor^{-1}$ (by causality), hence
$K \geq F\lfloor\frac{d_S}{a}\rfloor$, which implies that
$K \geq \log d_S + \log\l_{\max} + \log F - 2$.

The entanglement lower bound follows easily from conservation of entanglement
under local operations and classical communication (LOCC):
let Alice and Bob prepare a maximally entangled state $\ket{\Phi_0}$
of Schmidt rank $d_S$. If they were able to
do this exactly, by the non-increase of entanglement under LOCC, they
would need to start with at least $\log d_S$ ebits. However, the
protocol only succeeds in creating a state $\rho$ of fidelity
$\geq F$ with $\ket{\Phi_0}$. By a result of Nielsen~\cite{Nielsen00},
this implies that for the entanglement of formation,
\[
  E_F(\rho) 
  \geq \log d_S - 18\sqrt{1-F}\log d_S - 2\eta\left(2\sqrt{1-F}\right).
\]
Since $E_F$ cannot increase under LOCC, the right hand side is also
a lower bound on the number of ebits Alice and Bob started with.
\end{proof}

\begin{corollary}
A superdense coding protocol of fidelity $F\geq 1/2$
for all $d_S$-dimensional states with maximum
Schmidt coefficient $\leq\sqrt{\l_{\max}}$ must 
make use of at least
$\frac{1}{2}\log d_S + \frac{1}{2}\log\l_{\max} + \frac{1}{2}\log F - 1$ 
qubits of communication.
The sum of qubit and ebit resources must be at least
$\log d_S - 18\sqrt{1-F}\log d_S - 2\eta\left(2\sqrt{1-F}\right)$.
\end{corollary}
\begin{proof}
Suppose there exists an superdense coding  
protocol which can prepare all $d_S$ 
dimensional states with maximum Schmidt coefficient $\leq\sqrt{\l_{\max}}$
and which uses only $Q$ qubits and $E$ ebits.
Use teleportation to transmit the qubits, turning it 
into a remote state preparation protocol.

The qubit cost translates directly to a cbit cost of
$2Q$. From Proposition~\ref{prop:ersp} we infer the lower bound on $Q$.
The protocol including teleportation requires $Q+E$ ebits, thus
the lower bound on $Q+E$ follows from Proposition~\ref{prop:ersp}
as well.
\end{proof}

Thus, when $F\rightarrow 1$ and ignoring terms of order $o(\log d_S)$,
the upper resource bounds from our protocol, and the above lower
bound coincide.

\section{Protocol for a memoryless source} \label{sec:memoryless}

The universal protocol of Proposition~\ref{prop:esdc} is easily adapted to
the task of sending states produced by a memoryless source. A standard
application of typical subspace techniques gives control of the
value of $\l_{\max}$ and the effective size of the states received by Bob,
the two parameters determining the resources consumed by the universal 
protocol. We model the source 
$\cE_{A_1 S} = \{ p_i, \ket{\ph_i^{A_1 S}} \}_{i=1}^m$
as a sequence of independent, identically distributed states with 
\begin{equation}
\ket{\ph_{i^n}^{A_1 S}} 
= \ket{\ph^{A_1 S}_{i_1}}\ox \cdots \ox\ket{\ph^{A_1 S}_{i_n}}
\end{equation}
occuring with probability $p_{i^n} = p_{i_1}p_{i_2}\ldots p_{i_n}$,
where $i^n = i_1 i_2 \cdots i_n$.
If we define $S(\cE_S) = S\left(\sum_i p_i \Tr_{A_1} \proj{\ph_i}\right)$
and $\bar{S}(\cE_S) = \sum_i p_i S(\Tr_{A_1} \proj{\ph_i})$,
Harrow combined coherent classical communication and a remote state
preparation protocol to demonstrate that a qubit rate of 
$\frac{1}{2}(S(\cE_S) - \bar{S}(\cE_S))$ and ebit rate of
$\frac{1}{2}(S(\cE_S)+\bar{S}(\cE_S))$ are simultaneously 
achievable~\cite{H04}, an optimal result~\cite{AH03} which hinted at 
the existence of the universal protocol. Here we show how the 
universal protocol provides an alternate, perhaps more direct, 
route to Harrow's rate pair.

\begin{proposition} \label{prop:memorylessSDC}
There exist protocols for superdense coding of entangled states
with mean fidelity approaching one and
asymptotically achieving the rate pair of
$\frac{1}{2}(S(\cE_S) - \bar{S}(\cE_S))$ qubits and
$\frac{1}{2}(S(\cE_S)+\bar{S}(\cE_S))$ ebits.
\end{proposition}

\begin{proof}
With probability $p_{i^n}$, Alice needs to
prepare the state $\ket{\ph^{A_1 S}_{i^n}}$.
Instead, for typical $i^n$, she prepares a state $\ket{\s^{A_1 S}_{i^n}}$
obtained by applying a typical projector and a conditional
typical projector to $\ph^{S}_{i^n}$. When $i^n$ is atypical, the protocol
fails.

Given a probability distribution $q$ on a 
finite set $\chi$, define the set of \emph{typical sequences}, with
$\d > 0$, as
\begin{equation}
\cT_{q,\d}^n = \Big\{ x^n: \forall x | N(x|x^n) - nq_x | \leq 
                  \d \sqrt{n} \sqrt{q_x(1-q_x)} \Big\},
\end{equation}
where $N(x|x^n)$ counts the numbers of occurrences of $x$ in the
string $x^n = x_1 x_2 \cdots x_n$. If $\r = \sum_i p_i \ph_i^S$
has spectral decomposition $\sum_{j=1}^{d_S} R(j)\Pi_j$, we then 
define the typical projector to be
\begin{equation}
\Pi_{\r,\d}^n = \sum_{j^n \in \cT_{R,\d}^n} \Pi_{j_1} \ox \cdots \ox \Pi_{j_n}
\end{equation}
and the conditional typical projector to be
\begin{equation} 
  \Pi^n_{\ph^S, \d}(i^n) 
  = \bigotimes_{i=1}^{m} \Pi^{I_i}_{\ph_i, \d}, 
\end{equation}
where $ I_i= \{ j\in [n]: i_j =i \}$ and $\Pi^{I_i}_{\ph_i, \d}$ refers to the
typical projector in the tensor product of the systems $j\in I_i$.
In terms of these definitions, $\s^{A_1 S}_{i^n}$, the state
Alice prepares instead of $\ph_{i^n}^{A_1 S}$, is proportional to
\begin{equation}
     (\1_{A_1} \ox \Pi^n_{\rho,\d} \Pi^n_{\ph^S, \d}(i^n)) \ph^{A_1 S}_{i^n} 
      (\1_{A_1} \ox \Pi^n_{\ph^S, \d}(i^n)\Pi^n_{\rho,\d}),
\end{equation}

With respect to approximation, the relevant property of these operators 
is that, defining
\begin{equation}
\xi_{i^n} = \Pi_{\ph^S,\d}^n(i^n) \ph_{i^n}^S \Pi_{\ph^S,\d}^n(i^n),
\end{equation}
we have
\begin{eqnarray}
&& \Tr[ \Pi_{\r,\d}^n \Pi_{\ph^S,\d}^n(i^n) \ph_{i^n}^S 
     \Pi_{\ph^S,\d}^n(i^n) \Pi_{\r,\d}^n ] \nonumber \\
&& \phantom{==} 
   = \Tr[ \xi_{i^n} ] - \Tr[ (\1 - \Pi_{\r,\d}^n) \xi_{i^n} ]  \nonumber \\
&& \phantom{==} 
   \geq \Tr[ \xi_{i^n} ] - \Tr[ (\1 - \Pi_{\r,\d}^n) \ph_{i^n}^S ] \geq 1 - \e,
\end{eqnarray}
if $\d = m \sqrt{2d_S/\e}$ (by Lemmas 3 and 6 in~\cite{W99}).
The Gentle Measurement Lemma, referred to as the tender
operator inequality in~\cite{W99}, together with a simple
application of the triangle inequality implies that
$\| \ph_{i^n}^{A_1 S} - \s_{i^n}^{A_1 S} \|_1 \leq \sqrt{8\e} + 2\e$. 
For a more detailed proof of these
facts and further information about typical projectors, see~\cite{W99}.
If $i^n$ is typical, meaning it is in the set $\cT_{p,\d}^n$ (which
occurs with probability at least $1 - m/\d^2$), then it is also true
that
\begin{eqnarray}
 \label{eq:cond-typ}
 \Pi^n_{\ph^S, \d}(i^n) \ph^S_{i^n} \Pi^n_{\ph^S, \d}(i^n)
                      &\leq& \Pi^n_{\ph^S, \d}(i^n) 
                             2^{-n\bar{S}(\cE_S)+c\d\sqrt{n}}, \;\; \\
 \label{eq:typ}
 \Rank\Pi^n_{\rho,\d} &\leq& 2^{nS(\cE_S)+c\d\sqrt{n}},
\end{eqnarray}
where $c>0$ is independent of $n$ and $\d$. 
Equation~(\ref{eq:cond-typ}) implies that
$(1 - \e)\s^S_{i^n} \leq 2^{-n\bar{S}(\cE_S)+c\d\sqrt{n}}\Pi^n_{\rho,\d}$,
which in turn leads to the conclusion that 
$\l_{\max}(\s^S_{i^n}) \leq \smfrac{1}{1-\e}2^{-n\bar{S}(\cE_S)+c\d\sqrt{n}} 
=: \l_{\max}$;
Eq.~(\ref{eq:typ}) provides a bound on the effective dimension of
the system $S$ since $\s^S_{i^n} \leq \Pi^n_{\r,\d}$ for all $i^n$.

Applying the universal superdense coding 
protocol to $\s^{A_1 S}_{i^n}$, we find that
the number of qubits that must be sent is 
\begin{eqnarray}
&& \smfrac{1}{2}[ \log \Rank \Pi^n_{\r,\d} + \log \l_{\max} ] + o(n) \\
&& \quad 
   \leq \smfrac{n}{2}[S(\cE_S)-\bar{S}(\cE_S)] + c\d\sqrt{n} 
        - \log(1-\e) + o(n), \nonumber
\end{eqnarray}
while the number of ebits used is
\begin{eqnarray}
&& \smfrac{1}{2}[ \log \Rank \Pi^n_{\r,\d} - \log \l_{\max} ] + o(n) \\
&& \quad 
   \leq \smfrac{n}{2}[S(\cE_S)+\bar{S}(\cE_S)] + \log(1-\e) + o(n), \nonumber
\end{eqnarray}
matching the rates of the proposition.
\end{proof}

In Section \ref{sec:optimality} we used 
the fact that teleporting the qubits of a superdense
coding protocol leads to a remote state preparation protocol.
When applied to Proposition \ref{prop:memorylessSDC}, we get 
an alternative proof of Proposition 15 of \cite{BHLSW03}:
\begin{corollary}
There exist protocols for remote state preparation of entangled states
with mean fidelity approaching one and asymptotically achieving the
rate pair of $S(\cE_S) - \bar{S}(\cE_S)$ cbits and $S(\cE_S)$ ebits.
\qed
\end{corollary}

\section{Identification} \label{sec:ID}

Quantum message identification, a generalization of hypothesis testing to 
the quantum setting, has been explored recently in a series of 
papers~\cite{AW02,W04,W04.2}. As opposed to transmission, where the goal is to
reliably communicate a message over a channel, identification only allows
the receiver to answer a single binary question:
is the message $x$ or is it not?
A surprising aspect of the theory of identification
is that the number of questions that can be answered grows as a doubly 
exponential function of
the number of uses of the channel, as opposed to the well-known
singly exponential behavior for transmission~\cite{RY79,AD89}. In the quantum
setting, a number of versions of the identification (ID) capacity have been
defined; these divide broadly into the capacities for quantum resources
to identify classical messages and the capacities for those quantum resources
to identify quantum messages. In the former case, doubly exponential
growth of the number of messages was found, with the most important
result to date that the ID capacity of an ebit, supplemented with negligible
rate of forward classical communication, is two~\cite{W04.2}.
It follows, of course, that the ID capacity of a qubit is also two~\cite{W04}.

In this section, we will instead be focusing on the capacity of an ebit 
to identify quantum messages, that is, quantum states. We will consider
the model with a visible encoder and ID-visible decoder, according to
the terminology introduced in~\cite{W04}.

Specifically, we say that we have a 
quantum-ID code on $\cB(\CC^d) $ of error $0 < \l < 1$ and dimension 
$d_C$ if there exists an encoding map 
$\varepsilon:\cB(\CC^{d_C}) \rightarrow \cB(\CC^{d})$ and
a decoding map $D :\CC^{d_C} \rightarrow \cB(\CC^{d})$
such that for all pure states $\ket{\ph}$ and $\ket{\ps}$ on $\CC^{d_C}$
\begin{equation}
  \Bigl| \Tr(\ph \ps) - \Tr(\varepsilon(\ph)D_{\ps}) \Bigr| 
  \leq \frac{\l}{2}.
\end{equation}
This condition ensures that the measurement $(D_\ps,\1-D_\ps)$ can
be used on the states $\varepsilon(\ph)$ to simulate the test 
$(\ps,\1-\ps)$ applied to the states $\ph$.
In the blind encoder, ID-visible decoder case, 
$\varepsilon$ must be quantum channel and $D$ can be an arbitrary assignment
to operators $0 \leq D_\ps \leq \1$.
It was shown in \cite{W04} that for all $0 < \l < 1$ there exists a
constant $c(\l)>0$ such that on
$\CC^d$ a quantum-ID code of error $\l$ and 
$d_C = \lfloor c(\l) d^2 \rfloor$ exists.
Since, for fixed $\l$, $\log d_C = 2 \log d - \mbox{const}$, this
shows that, asymptotically, one qubit of communication can identify two
qubits. We claim that, again asymptotically, but now using a visible 
encoding map,
one ebit plus a negligible (rate of) quantum communication can be used
to identify two qubits. Rather than providing a detailed argument,
we simply state the method: the states $\varepsilon(\ph)$ 
that are output by the blind encoding can be prepared visibly using superdense
coding. Because they are extremely mixed, their purifications are highly
entangled and Proposition \ref{prop:esdc} demonstrates that negligible
communication is sufficient. 

The negligible communication cost is encountered frequently in the
theory of identification: the classical identification capacity of a bit
of shared randomness supplemented by negligible communication is a bit.
In \cite{W04.2}, it was found that the classical identification capacity of
an ebit supplemented by negligible communication is two bits.
Our finding here that the quantum identification capacity of an ebit and 
negligible communication is two qubits provides an alternative proof of this result.

\begin{proposition}
If $d_C = \left\lfloor c(\l) d^2 / (\log d)^4 \right\rfloor$, then for all states 
$\ket{\ph} \in \CC^{d_C}$,
approximations $\ket{\Phi_\ph'}$ of the purifications of the states 
$\varepsilon(\ph)$ can be prepared on $\CC^a \ox \CC^d$ using
$\log d + o(\log d)$ ebits and $o(\log d)$ qubits of communication, in
such a way that 
\begin{equation}
\Bigl | \Tr[\ph \ps] - \Tr[(\Tr_{\CC^a} \Phi_\ph')D_{\ps}] \Bigr | 
\leq \frac{\l}{2}.
\end{equation}
\end{proposition}
\begin{proof} 
From the proof of Proposition 17 in \cite{W04}, if we choose
$a =\lfloor \e d/2 \rfloor$, $\e = (\lambda/48)^2$, and
a $\lambda/16$-net in $\CC^{d_C}$, we may
let $\varepsilon(\ph) = \Tr_{\CC^a}(V\ph V^\dg)$
with $V: C^{d_C} \rightarrow \CC^d \ox \CC^a$ a Haar
distributed isometry, and
$D_{\ps} = \supp \, \varepsilon(\tilde\ps)$ for
the state $\tilde\ps$ closest to $\ps$ in the net. Then,
\begin{eqnarray}
 & & \Pr\Biggl( \exists \ps, \ph \mbox{ such that } 
         | \Tr(\ph \ps) - \Tr(\varepsilon(\ph)D_{\ps}) | 
         > \frac{\l}{4} \Biggr) \nonumber \\
 & & \quad \quad \quad \quad \quad \leq 
     \left(\frac{c_0}{\lambda}\right)^{4d_C} \exp(-c_1 d^2 \e^2),
\end{eqnarray}
with absolute constants $c_0$ and $c_1$. (Note that this statement is
trivial for $a=0$ or $a=1$.) To be precise, in~\cite{W04} the above
probability bound is derived for states in the net, but it is
also explained how to use triangle inequality to lift this
to all states.

Therefore, the states $\varepsilon(\ph)$ form a good quantum-ID code. We will 
demonstrate how to make them using superdense coding. Arguing along the lines 
of Eq.~(\ref{eq:complicatedthing}), 
we find that for all $\a > 0$, $d_C = \lfloor ad / (\log a)^4 \rfloor$ 
and sufficiently large $d$,
\begin{equation}
  \Pr\left( \underset{\ph}{\inf}\, S\bigl(\varepsilon(\ph)\bigr)
         < \log a - \a \right) < \frac{1}{2}.
\end{equation}
(This is also a special case of Theorem IV.1 from \cite{HLW04}.)
By the same reasoning given after Eq.~(\ref{Eq:traceEq}),
there exists a maximally entangled state $\ket{\Phi_\ph}$ such 
that $|\bra{\Phi_\ph}V\ket{\ph}|^2 \geq 1 - \sqrt{2\a\ln 2}$.
We can, therefore, invoke Proposition \ref{prop:esdc} with $d_S = d$
and $\l_{\max} = 1 / a$ to conclude that for
sufficiently large $d$, states $\ket{\Phi_\ph'}$ approximating
$\ket{\Phi_\ph}$ to within fidelity $\sqrt{2\a \ln 2}$
can be prepared
using $\log d + o(\log d)$ ebits and $o(\log d)$ qubits of communication.
By an appropriate choice of $\a$, we can therefore ensure that
$|\bra{\Phi_\ph}V\ket{\ph}|^2 \geq \l / 4$.
Using the triangle inequality, we then find that
\begin{equation}
|\Tr(\ph \ps) - \Tr(\Phi_\ph' D_\ps) | \leq \l/2
\end{equation}
for all pure states $\ket{\ph} \in \CC^{d_C}$.
\end{proof}

There is a little subtlety in the proof that is worth considering 
briefly. The states to be prepared, $\ket{\Phi_\ph}$, are maximally
entangled, so one might think that they can be prepared without
any communication at all. The party holding $\CC^d$ can, indeed, create
them without communication. The party holding the smaller $\CC^a$,
however, cannot; local unitary transformations on $\CC^a$ will not change
the support of the reduction to $\CC^d$, for example. Nonetheless, 
by appealing to Proposition \ref{prop:esdc}, we see that the asymmetry
disappears in the asymptotic limit if negligible communication
is allowed.

\section{Discussion}
We have proved the existence of protocols which allow a sender to share
entangled states with a receiver
while using as little quantum communication as is possible.
These protocols interpolate between requiring no communication at
all for maximally entangled states and a rate of two remote qubits
per sent qubit for product states.
An immediate application of the result was a proof that the identification
capacity of an ebit is two qubits when visible encoding is permitted.

The question of efficient constructions remains -- 
we would like to have protocols with the same ebit and qubit rates which
are implementable in polynomial time
(as has been demonstrated for state randomization~\cite{HLSW03}
by Ambainis and Smith~\cite{AS04}). It would also be interesting to
know
whether stronger success criteria can be satisfied while still achieving
the same rates. Specifically, the universal remote state preparation
protocol of \cite{BHLSW03} produces an \emph{exact} copy of 
the desired state when
the protocol succeeds, not just a high fidelity copy. Is such a 
probabilistic-exact protocol possible in the superdense coding setting?
(One could even ask questions about perfectly faithful 
superdense coding, in analogy
to what has been done for remote state preparation in \cite{P01,YZG04,B04}.)
Another
natural question is the quantum identification capacity of an
ebit in the \emph{blind} scenario. We have shown
that it is possible to achieve the identification rate of two qubits
per ebit in the case when the identity of the encoded qubits is known, 
but it is not at all clear 
whether this rate is achievable when the identity of the qubits is unknown.

\medskip

\subsection*{Acknowledgments}
It is a pleasure to thank Dave Bacon and Debbie Leung 
for many enjoyable discussions on subjects directly and peripherally related
to the contents of this paper.



\end{document}